\documentclass[reprint,amsmath,amssymb,aps,superscriptaddress]{revtex4-1}
\usepackage{graphicx,xspace,epstopdf,attachfile}

\begin{document}

\title{Diffusion-dominated pinch-off of ultralow surface tension fluids}

\author{Hau Yung Lo}
\affiliation{Department of Physics, The Chinese University of Hong Kong, Hong Kong, P. R. China}
\author{Yuan Liu}
\affiliation{Department of Physics, The Chinese University of Hong Kong, Hong Kong, P. R. China}
\affiliation{Department of Mechanical Engineering, University of Hong Kong, Hong Kong, P. R. China}
\author{Sze Yi Mak}
\affiliation{Department of Mechanical Engineering, University of Hong Kong, Hong Kong, P. R. China}
\author{Zhuo Xu}
\affiliation{Department of Physics, The Chinese University of Hong Kong, Hong Kong, P. R. China}
\author{Youchuang Chao}
\affiliation{Department of Mechanical Engineering, University of Hong Kong, Hong Kong, P. R. China}
\author{Kaye Jiale Li}
\affiliation{Department of Physics, The Chinese University of Hong Kong, Hong Kong, P. R. China}
\author{Ho Cheung Shum}
\email{ashum@hku.hk}
\affiliation{Department of Mechanical Engineering, University of Hong Kong, Hong Kong, P. R. China}
\author{Lei Xu}
\email{xuleixu@cuhk.edu.hk}
\affiliation{Department of Physics, The Chinese University of Hong Kong, Hong Kong, P. R. China}

\date{\today}

\begin{abstract}
We study the breakup of a liquid thread inside another liquid at different surface tensions. In general, the pinch-off of a liquid thread is governed by the dynamics of fluid flow. However, when the interfacial tension is ultralow (2 to 3 orders lower than normal liquids), we find that the pinch-off dynamics can be governed by bulk diffusion. By studying the velocity and the profile of the pinch-off, we explain why the diffusion-dominated pinch-off takes over the conventional breakup at ultralow surface tensions.
\end{abstract}

\maketitle
The breakup of a liquid thread, or pinch-off, has been extensively studied. The pinch-off dynamics are classified into different regimes according to its governing fluid mechanics \cite{ListerStone,regime1,regime2,regime3,regime4}. Typically, apart from a few exceptions \cite{NagelSci,AirPRL1,AirPRL2,AirJFM1,AirPRL3,NagelNP}, the thinning dynamics is driven by surface tension, and balanced by inertia and/or viscous resistance, and the interface motion is described by self-similar solutions which are independent of initial conditions \cite{review1,OneFluid1,inert1,TwoFluid1,TwoFluid2,OneFluid2,Bubble1,inert2,microfluid1,inert3,surf1,complex1,complex2,microfluid2,surf2}.

In the conventional theory, lowering the surface tension would slow down the pinch-off process, but the fundamental picture described by force balance and fluid flow remains essentially unchanged. However, recent studies find that when surface tension becomes extremely low, in the order of nN/m, the pinch-off dynamics would enter a new regime, which is governed by thermal fluctuation \cite{thermal1,thermal2}. Surprisingly, our experiment now reveals that, in addition to the thermal fluctuation regime, the pinch-off dynamics could also be dominated by bulk diffusion. This bulk diffusion regime appears when the interfacial tension is 2 to 3 orders lower than normal liquids (i.e., 0.1 to 0.01mN/m), but still much higher than that required by the thermal fluctuation regime.

The diffusion-dominated pinch-off represents a new class of breakup phenomenon, which was recently discovered in metal alloy at the liquid-solid interface \cite{Aagesen}. We now observe this phenomenon at the liquid-liquid interface for the first time, and identify a new regime for pinch-off dynamics inside fluid systems. Besides improving the general understanding of fluids pinch-off, our study could also make an important impact to the research field of breakup at very small length scales, such as the study of nanojets\cite{nano1,nano2}, and the area related to ultralow interfacial tensions, for instance the division of active droplets\cite{active1,active2} and the applications of aqueous two-phase systems \cite{ATPS1,ATPS2,ATPS3}.

Our experiment is performed in an aqueous two-phase system (ATPS) because its surface tension can be widely adjusted across 3 order of magnitudes. The ATPS we are using is a ternary mixture made of water, polyethylene glycol (PEG) and tripotassium phosphate \cite{component1,component2}. Two phases, one polymer-rich and one salt-rich, can spontaneously form and create a clear interface. By adjusting the ratio of the three components, the system can approach to or depart from a critical point at room temperature. Surface tension decreases significantly as the system approaches the critical point. In other words, by adjusting the concentration of the three components, systems with different surface tensions can be realized. To promote general validity, in addition to varying surface tensions, two types of PEG with distinct polymer chain lengths are used. We name these two groups of samples as ATPS-A and ATPS-B. A very small amount of fluorescent dye, 0.03wt\% Rhodamine-B, is added into the solution to increase the contrast between the polymer-rich phase and the salt-rich phase. The two phases are homogenous and have reached an equilibrium state. Surface tensions are measured by the spinning drop method \cite{measure1}. More details are provided in the supplementary information \cite{SupInfo}.

The pinch-off experiments at ultralow surface tensions are carried out in a thin cell of $50\mu m$ thick. The experiments at normal surface tensions are carried out in a flow-focusing microfluidic device (cross section: ${50\times100 \mu m}$), because pinch-off events at normal surface tensions rarely occur in the thin cell. We inject one phase of liquid into another phase, and then observe the narrowing liquid thread and pinch-off event by a fluorescence microscope coupled with a high-speed camera, with $63\times$ magnification and 5,000-10,000 frames per second. We make sure that the narrowing liquid thread locates near the middle of the cell without touching any boundary walls.

\begin{figure*}
\includegraphics[width=172mm]{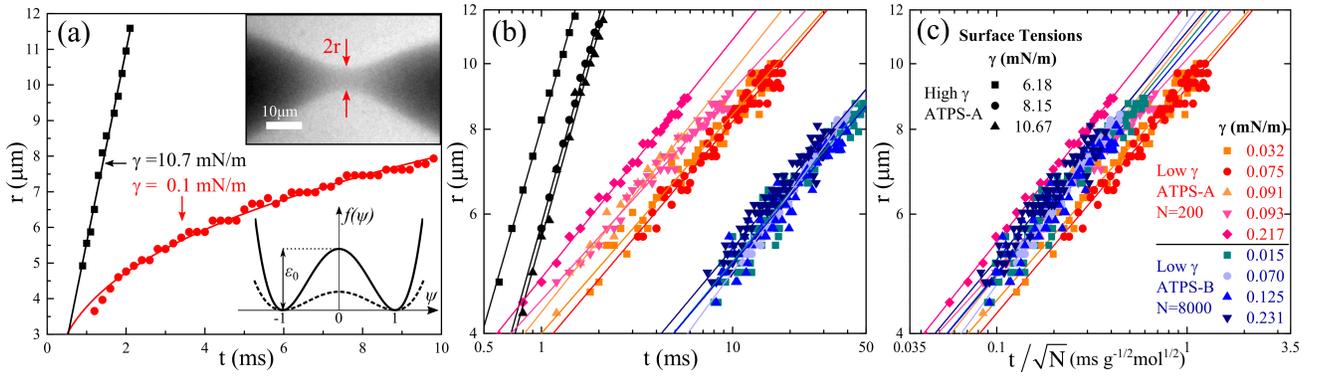}
\caption{\label{figM1}
Neck radius $r$ as a function of the time to pinch-off ${t}$. \textbf{a)} Two example curves in which there is a 100 times difference in surface tension ${\gamma}$. The solid curves are the best fit. For ${\gamma}$ = 10.7 mN/m, the curve is linear. For ${\gamma}$ = 0.1 mN/m, the curve follows a power law with an exponent ${0.33}$. Both samples are ATPS-A samples. The upper inset shows a fluorescence image of liquid thread before pinch-off. The outer phase (polymer-rich) is brighter than the inner phase (salt-rich). The scale bar is ${10 \mu m}$. The lower inset shows a sketch of free energy (see Eq.~(\ref{eqM5})) and the dashed curve indicates a near-critical situation with a very small $\varepsilon_0$.
\textbf{b)} Log-log plots of typical measurements for samples with various surface tensions. The solid lines are the best fit.
\textbf{c)} Replotting the low surface tension data in b) as a function of $ t/\sqrt{N}$, where $N$ are the molecular weights of polymers. b) and c) share the same legend.
}
\end{figure*}

We first demonstrate typical pinch-off events in Fig.~\ref{figM1} (a) and Movie 1. The narrowing process of the liquid thread is measured by a high speed camera as shown in Movie 1. In Fig.~\ref{figM1} (a), the neck radius measured at the thinnest location, $r$, is plotted against the time to pinch-off, ${t}$. Here we define ${t = 0}$ as the moment of pinch-off. To make a clear comparison, we choose two systems with 100 times difference in surface tension. The higher system has $\gamma = 10.7 mN/m$, comparable to that between olive oil and water. For this normal system, the neck radius decreases linearly with time, as expected in the Stokes regime. By contrast, the lower system has an ultralow surface tension, $\gamma = 0.1mN/m$, and the neck radius follows a power law with an exponent ${n = 0.33}$. Such a significant change in the narrowing dynamic clearly indicates a new pinch-off regime. To make sure its generality, we repeat the same measurement on various samples with different surface tensions and viscosities as shown in Fig.~\ref{figM1} (b). Clearly there exist two distinct classes of behaviors.

Next, we theoretically show that the new behavior at low surface tensions is due to bulk diffusion, and this diffusion regime must dominate the Stokes regime at sufficiently low surface tensions. For viscous liquids, pinch-off is governed by the balance between the capillary pressure from surface tension and the viscous resistance from viscosity, leading to the famous Stokes regime \cite{ListerStone,TwoFluid1,TwoFluid2}. In this regime, the neck radius of the liquid thread, ${r_{St}}$, is a linear function of time:
\begin{equation}\label{eqM1}
r_{St}=h(m)\cdot \gamma / \mu \cdot t
\end{equation}
\noindent where ${t}$ is the time to pinch-off, ${\gamma}$ is surface tension, ${\mu}$ and ${m \mu}$ are viscosities of the liquid thread (inner phase) and surrounding liquid (outer phase) respectively, and ${h}$ is a dimensionless prefactor depending only on $m$. The corresponding radial velocity is simply a constant:
\begin{equation}\label{eqM2}
v_{St} \equiv \frac{d}{dt}r_{St} = h(m)\cdot \gamma / \mu
\end{equation}

For the bulk diffusion regime, the underlying mechanism is entirely different. In the model established by Aagesen \emph{et al} \cite{Aagesen}, the curvature of the interface changes the equilibrium concentration at the interface, as described by Gibbs-Thomson equation \cite{VoorReview,BookVoor}, which subsequently induces bulk diffusion and hence movement of the interface. This movement is governed by diffusion equations coupled with boundary conditions at the interface, and forms a new class of self-similar interface motion. In this regime, the neck radius follows a one-third power law with respect to time, consistent with our observation:
\begin{equation}\label{eqM3}
r_\mathit{diff} = f_{0}(Dd_{0}t)^{1/3}
\end{equation}
\noindent where ${f_{0}}$ is a dimensionless constant, $D$ is the diffusivity in the inner phase, and ${d_{0}}$ is the capillary length defined thermodynamically \cite{VoorReview,BookVoor} (not $\sqrt{\gamma/\Delta\rho g}$). The corresponding radial velocity is:
\begin{equation}\label{eqM4}
v_\mathit{diff} \equiv \frac{d}{dt}r_\mathit{diff} = \frac{1}{3} \frac{f^{3}_{0} D d_{0}}{r^{2}_\mathit{diff}}
\end{equation}

To summarize, the thinning dynamics in Stoke regime is governed by mechanical equilibrium, while in diffusion regime it is governed by chemical equilibrium. In general, mechanical and chemical mechanisms should coexist and be independent of each other; however, in real situations we should only be able to observe the dynamics driven by the dominant factor. Here we hypothesize that the dominant mechanism would simply be the one which causes a faster pinch-off. At normal surface tensions, the mechanical driving force moves the interface so fast that the diffusion effect can be totally neglected; however, at ultralow surface tensions, the mechanical driving force reduces significantly and the diffusion effect catches up and takes over. This hypothesis will be further tested by comparing relevant speeds caused by the two mechanisms, both theoretically and experimentally.

To theoretically compare the speeds of the two mechanisms, ${v_\mathit{diff}}$ and ${v_{St}}$, we need to relate ${v_\mathit{diff}}$ to liquid properties, $\gamma$ and $\mu$, as the following. Let the Landau free energy density around the critical point be \cite{BookDesai}
\begin{equation}\label{eqM5}
f = \varepsilon_{0}(\psi^{4} - 2 \psi^{2} +1)
\end{equation}
\noindent where ${\varepsilon_{0} \equiv f(0)}$ is the energy barrier between the two phases in our system, ${\psi}$ is the normalized concentration such that $f$ has two minima at ${\psi = \pm1}$, which correspond to the two phases of our system. The profile of the free energy is illustrated in the lower inset of Fig.~\ref{figM1}(a). One can show that $d_0$ is related to $\gamma$ (see SI) \cite{BookDesai,BookChaikin,SupInfo}
\begin{equation}\label{eqM6}
\frac{\gamma}{d_{0}} = 16 \varepsilon_{0}
\end{equation}
\noindent Furthermore, by the Zimm model and Stokes-Einstein relation, the diffusivity is ${D = k_{B}T / (6 \pi \mu_{s} a)}$, where ${k_{B}T}$ is the product of Boltzmann constant and temperature, ${a\propto \sqrt{N}}$ is the hydrodynamic radius of polymers, $N$ is the molecular weight, and $\mu_{s}$ is the viscosity of the solvent, which is close to the solution viscosity $\mu$ in our samples (see SI) \cite{SupInfo}. Compared with surface tension, the viscosity remains roughly a constant as the critical point is approached. Therefore, Eq.~(\ref{eqM4}) can be expressed as
\begin{equation}\label{eqM7}
v_\mathit{diff} \propto \frac{k_{B}T}{a r^{2}_\mathit{diff}}\frac{\gamma}{\mu_s}\frac{1}{\varepsilon_{0}}  \propto \frac{k_{B}T}{\sqrt{N} r^{2}_\mathit{diff}}\frac{\gamma}{\mu_s}\frac{1}{\varepsilon_{0}}
\end{equation}
\noindent Finally, by Eq.~(\ref{eqM2}),(\ref{eqM7}), at constant temperature and neck radius, we have
 \begin{equation}\label{eqM8}
\frac{v_\mathit{diff}}{v_{St}}  \propto \frac{1}{\varepsilon_{0}}
\end{equation}
\noindent Because the energy barrier ${\varepsilon_{0}}$ would vanish at the critical point, therefore, Eq.~(\ref{eqM8}) gives ${v_\mathit{diff} \gg v_{St}}$ near the critical point. At the same time, the surface tension also approaches zero near the critical point. Consequently, the bulk diffusion regime would eventually dominate the Stoke regime when the surface tension is sufficiently low, or equivalently, when the system is sufficiently close to the critical point.

Note that the model assumes a binary system, but a ternary mixture of polymer, salt and water is used in our experiment. Nevertheless, we argue that our system can be `coarse-grained' as a binary system by treating salt and water as the high-diffusivity component, and polymer as the low-diffusivity component.

\begin{figure}
\includegraphics[width=86mm]{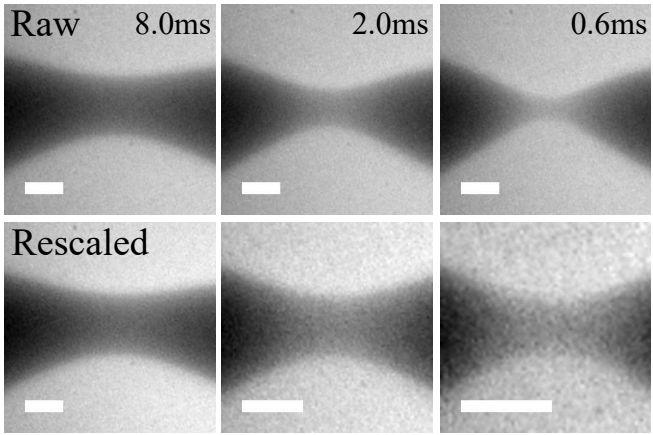}
\caption{\label{figMmovie}
Snapshots of movie 1 showing the self-similarity of the interface. Top row: Raw images of the pinch-off process at different time. Bottom Row: Corresponding rescaled images, which are resacled by a length scale ${L \equiv t^{n}}$ in both radial and axial directions. The exponent $n$ is obtained from fitting the $r$-$t$ curves as shown in Fig.~\ref{figM1}. The scale bars in all images are 10$\mu$m.
}
\end{figure}

To confirm the existence of the bulk diffusion regime in our system, we carefully compare the experimental results with the established model predictions \cite{Aagesen}. The model predicts several characteristic features: First, the interface profile should be self-similar with the same time-dependent length scale in both radial and axial directions. Fig.~\ref{figMmovie} and Movie 1 confirm this self-similarity by rescaling the captured images with \emph{the same} length scale, ${L \equiv t^{n}}$, in both radial and axial directions. The exponent $n$ is obtained from fitting the $r$-$t$ curves as shown in Fig.~\ref{figM1}.

\begin{figure}
\includegraphics[width=86mm]{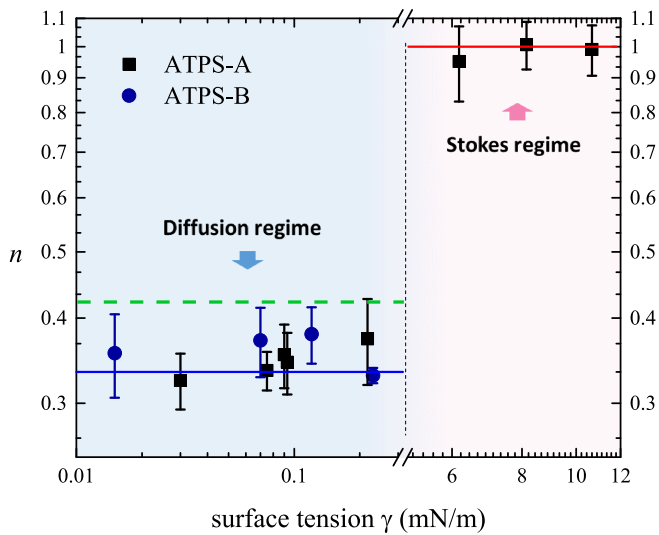}
\caption{\label{figM2}
Mean exponents $n$ for different surface tensions ${\gamma}$ in log-log scale. The error bars include standard deviations of multiple measurements and instrumental uncertainties. The red line indicates the characteristic exponent of Stokes regime, which is 1. The blue line indicates the characteristic exponent of bulk diffusion regime, which is 1/3. The measured exponents are consistent with these two corresponding characteristic exponents. The green dashed line indicates the characteristic exponent of thermal fluctuation regime(${\sim 0.42}$). Apparently it does not agree with the data.}
\end{figure}

Second, the neck radius should follow ${r \propto t^{1/3}}$ as mentioned in Eq.~(\ref{eqM3}). In Fig.~\ref{figM2}, we plot the exponents measured experimentally across a broad range of surface tensions. The error bars include standard deviation of multiple measurements and instrumental uncertainties. At high surface tensions, the exponents agree well with the one in Stokes regime $n=1$ as expected (red line). At low surface tensions, the exponents are very different: across a wide $\gamma$ range from 0.015 to 0.2 ${mN/m}$, the exponents lie between 0.32 and 0.38. It is consistent with the characteristic exponent of diffusion regime, $n=1/3$ (blue line). Note that there is a break in the horizontal axis, corresponding to the intermediate surface tension range (0.3 to 6${mN/m}$). The experimental results in this range have very low reproducibility, indicating complicated dynamics in this transition range, where two factors with similar magnitudes compete with each other. Because of low reproducibility, we do not discuss this range in the paper.

Third, the Stokes-Einstein relation and Zimm model predict that the diffusivity scales with the molecular weight, $D\propto 1/\sqrt{N}$ (see Eq.~(\ref{eqM7}). Therefore, the pinch-off dynamics of APTS-A and APTS-B should also scale with $N$: $ r_\mathit{diff} \propto (Dt)^{1/3} \propto(t/\sqrt{N})^{1/3}$. To verify it, we plot the rescaled data in Fig.~\ref{figM1}(c): clearly the two separated sets of data with 1/3 power in Fig.~\ref{figM1}(b) collapse excellently in Fig.~\ref{figM1}(c). This nice collapse once again indicates that our pinch-off dynamics is controlled by diffusion.

\begin{figure}
\includegraphics[width=86mm]{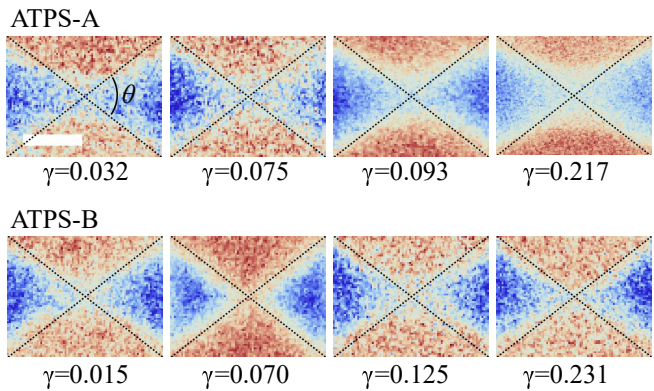}
\caption{\label{figM3}
Overlaying the theoretical cone angle on the captured images at different surface tensions at $t=1$ ms. The black dashed lines represent the theoretical asymptotic cone angle of ${\theta = 76^{o}}$.
The brightness of the captured images are displayed in pseudo-color to enhance contrast. The blue region is the inner phase, the red region is the outer phase. The scale bar is ${10 \mu m}$.}
\end{figure}

Fourth, the profile of interface should be symmetric with an asymptotic cone angle of ${76^{o}}$. By contrast, the profile in the Stokes regime is asymmetric \cite{TwoFluid1,TwoFluid2}. In Fig.~\ref{figM3}, we overlay the theoretical ${76^{o}}$ cone (black dashed line) on the captured images for different low-surface-tension experiments, at $t=$1ms before pinch-off. The experimental angle is consistent with the theoretical value, with a small deviation around several degrees. This could be caused by the non-zero diffusivity of the outer phase (polymer-rich phase) in experiment, which is inconsistent with the liquid-solid model. It is reasonable that the corresponding error in cone angle is small, because even in the extreme case where the liquid and solid phases are interchanged, the cone angle only changes from ${76^{o}}$ to ${80^{o}}$ \cite{Aagesen}.

Next, we need to exclude the possibility that the observed pinch-off is governed by thermal fluctuation, which also occurs at low surface tensions. First, for the thermal fluctuation regime, the characteristic exponent is 0.42 \cite{thermal1,thermal2} and the characteristic cone angle is ${\sim 30^{o}}$\cite{thermal1}, which are inconsistent with our measurements shown in Fig.~\ref{figM2} and Fig.~\ref{figM3}. Second, it is well known that thermal fluctuation will dominate only when the size of the system is comparable to the thermal length scale, ${L_{th} = \sqrt{kT/\gamma}}$ \cite{thermal1,thermal2}. In our system, to obtain ${L_{th} \sim 1 \mu m}$, the surface tension needs to reach ${nN/m}$, which is ${10^{4}}$ times smaller than the lowest value in our experiment. Therefore, our results represent a new regime irrelevant to thermal fluctuation.

We also discuss the possibility of viscoelastic effect in our polymer solutions. By measuring shear stress versus shear rate, we confirm that our experimental systems are Newtonian (see SI)\cite{SupInfo}. In addition, the viscoelastic effect typically induces an exponential dependence of $r$ versus $t$, while our results are well described by a 1/3 power law \cite{complex1,complex2}. Moreover, previous studies have shown that varying polymer concentrations and chain lengths induce significant changes in the viscoelastic response \cite{polymer}, while our pinch-off behaviors under such various conditions collapse onto a universal power law (see Fig.1c). Therefore, we believe that the viscoelastic effect does not play a significant role.

\begin{figure}
\includegraphics[width=86mm]{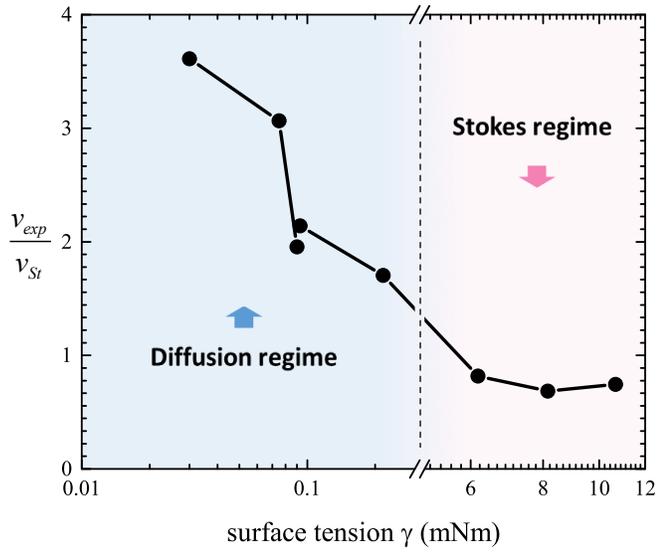}
\caption{\label{figM5}
The ratio between measured velocity $v_{exp}$ and theoretical velocity of Stoke regime ${v_{St}}$. The theoretical velocity ${v_{St}}$ is calculated without any fitting parameter. The velocities are measured or calculated at ${r = 5\mu m}$. All data points correspond to ATPS-A samples. The large values in the diffusion regime quantitatively verify that ${v_\mathit{diff} \gg v_{St}}$ and the faster mechanism dominates the final outcome.
}
\end{figure}

We have shown that, by Eq.~(\ref{eqM7}-\ref{eqM8}), the velocity of diffusion regime would surpass that of Stokes regime when surface tension is sufficiently low, and hence become dominant. To verify, we compare the measured velocities $v_{exp}$ with the calculated ${v_{St}}$ by using Eq.~(\ref{eqM2}) without any fitting parameter, as shown in Fig.~\ref{figM5}. On the right side in the Stokes regime, the ratios are close to unity ($\sim$0.8) as expected. This proves that our calculations are reliable. On the left side in the diffusion regime, the measured velocities are several times faster than ${v_{St}}$. This confirms our hypothesis that, while both the mechanical and the chemical pinch-off mechanisms coexist, only the faster one will materialize. The origin of ${v_\mathit{diff}}$ takes over ${v_{St}}$ comes from the relation that, as predicted by Eq.~(\ref{eqM7}), ${v_\mathit{diff}/\gamma}$ increases substantially as surface tension decreases. The corresponding data are provided in the supplementary information \cite{SupInfo}. Note that in the calculations of ${v_{St}}$, we adopt the dimensionless prefactor of ${h(m) = 0.0335 \cdot m^{-0.47}}$ given by Cohen \emph{et al} \cite{TwoFluid1}. This expression of ${h(m)}$ is valid for ${m \sim 1}$ (i.e., two liquids with similar viscosities), and thus only the ATPS-A samples fulfill this requirement and are plotted. Same analysis for ATPS-B shows a similar trend, as shown in Fig.S6 in the supplementary information \cite{SupInfo}.

To conclude, we study the pinch-off process in two-fluid systems across a wide range of surface tensions. We find a new regime governed by bulk diffusion at ultralow interfacial tensions, which locates below the normal Stokes regime and above the thermal fluctuation regime. We verify this diffusion regime by comparing the measurements with the theoretical predictions. We further hypothesize and verify that when both mechanical and diffusion mechanisms coexist, only the much faster one will materialize. Even beyond the ultralow surface tension systems, we suspect that the diffusion regime may generally appear at small enough neck radius (see Eq.~(\ref{eqM7})), and suggest more studies in this direction.

\begin{acknowledgments}
This project is supported by Hong Kong RGC (GRF 14306518, 14303415, 17304017, 17305414, CRF C6004-14G, C1018-17G), and CUHK Direct Grants (4053313, 4053231, 4053167, 4053354). We thank Peter W. Voorhees, Xiaohu Zhou, Yang Song and Hao Yuan for helpful discussions. We also acknowledge an anonymous referee for suggesting the Zimm model.

H.Y.Lo and Y.Liu contributed equally to this work.
\end{acknowledgments}

\newpage
\newpage

\newpage

\newpage

\newpage

\newpage

\end{document}